\documentclass[9pt,twocolumn,twoside]{pnas-new}
\setboolean{displaywatermark}{false}
\templatetype{pnasresearcharticle} % Choose template
\title{The Nernst Effect in Corbino Geometry}

\author[a,1,2]{A.V.~Kavokin}
\author[a,b,1]{B.L.~Altshuler}
\author[c,1]{S.G.~Sharapov}
\author[d,1]{P.S.~Grigoryev}
\author[e,1]{A.A.~Varlamov}

\affil[a]{Westlake University, 18 Shilongshan Road, Hangzhou 310024, Zhejiang Province, China}
\affil[b]{Department of Physics, Columbia University, NY, USA}
\affil[c]{Bogolyubov Institute for Theoretical Physics, National Academy of
Science of Ukraine, 14-b Metrologichna Street, Kyiv, 03680, Ukraine}
\affil[d]{Spin Optics Laboratory, St. Petersburg State University, Ulyanovskaya 1, 198504 St. Petersburg, Russia}
\affil[e]{CNR-SPIN, Viale del Politecnico 1, I-00133, Rome, Italy}

\leadauthor{Kavokin}

\significancestatement{The Nernst effect consists in the induction of an electric current by a combined effect of the external magnetic field and the temperature gradient. We consider a Corbino disk geometry, where the temperature difference is applied between the outer and inner edges of the disk, while the magnetic field is perpendicular to the plane of the disk. We show that the circular diamagnetic currents flowing along the edges of the disk are the oscillatory functions of the filling factor of Landau levels of the electron gas in the disk. The Corbino geometry offers a unique opportunity for observation of the magnetization currents that have a purely thermodynamic nature, e.g. are independent of the conductivity of the sample.}

\equalauthors{\textsuperscript{1} All the authors contributed equally to this work and to writing the manuscript.
}
\correspondingauthor{\textsuperscript{2} E-mail: a.kavokin@westlake.edu.cn}


\begin{abstract}
We study the manifestation of the  Nernst effect in the Corbino disk subjected to the normal external magnetic field and to the radial temperature gradient. The Corbino geometry offers a precious opportunity for the direct measurement of the magnetization currents that are masked by kinetic contributions to the Nernst current in the conventional geometry. The magnetization currents, also referred to as the edge currents, are independent on the conductivity of the sample which is why they can be conveniently described within the thermodynamic approach. They can be related to the Landau thermodynamic potential for an infinite system. We  demonstrate  that
the observable manifestation of this, purely thermodynamic, Nernst effect consists in the strong oscillations of the  magnetic field measured in the center of the disk as a function of the external field. The oscillations depend on the temperature difference at the edges of the disk. Dirac fermions and 2D electrons with a parabolic spectrum are characterized by oscillations of different phase and frequency. We predict qualitatively different power dependencies of the magnitude of the Nernst signal on the chemical potential for normal and Dirac carriers. 
\end{abstract} 

\dates{This manuscript was compiled on \today}
\doi{\url{www.pnas.org/cgi/doi/10.1073/pnas.XXXXXXXXXX}}

\begin{document}

\maketitle
\thispagestyle{firststyle}
\ifthenelse{\boolean{shortarticle}}{\ifthenelse{\boolean{singlecolumn}}{\abscontentformatted}{\abscontent}}{}

% If your first paragraph (i.e. with the \dropcap) contains a list environment (quote, quotation, theorem, definition, enumerate, itemize...), the line after the list may have some extra indentation. If this is the case, add \parshape=0 to the end of the list environment.
\dropcap{A} Corbino disk represents one of the most important experimental designs
for studies of transport effects in solids \cite{Corbino1911}. In contrast
to the Hall bar geometry \cite{Hajdu1994book}, in a Corbino disk the Lorentz
force induced by a magnetic field normal to the plane of the structure
is not compensated by the induced electrostatic force. The Lorentz
force gives rise to circular edge currents that can be studied through
the magnetization generated by them \cite{Laughlin1981PRB}. These currents
are usually referred to as magnetization or diamagnetic currents \cite{Obraztsov1965}. They are governed by the gradient of the magnetisation in a sample and are formally independent of the electric conductivity henceforth. 
In classical language they arise because of the reflection of carriers circulating on
their cyclotron orbits from the inner and the outer edges of the disk \cite{Teller1931,Hajdu1974ZP,Mineev2007PRB}.
In the range of classically strong magnetic fields, the magnetization currents
exhibit oscillations with a periodicity governed by the resonances
between Fermi and Landau energy levels \cite{LKV2011PRL}. These oscillations can be studied e.g. by measuring the magnetic field induced by edge currents in the center of the disk.

The Hall effect in the Corbino geometry has been studied both in classical \cite{Adams1915}
and quantum \cite{Dolgopolov1992PRB} limits. In contrast, the most
important thermomagnetic effect, namely the Nernst effect, remains
poorly explored in the disk geometry. The Nernst effect \cite{Nernst1886}
consists in the induction of an electric current by a combined action
of the crossed external magnetic field and the temperature gradient.
It may be considered as a heat counterpart of the Hall effect. Recently,
the giant Nernst or Nernst-Ettingshausen effects have been observed
in graphene \cite{Zuev2009PRL,Chekelsky2009PRB}, in pseudogap phase of quasi-two dimensional high temperature superconductors
\cite{Palstra1990PRL,Zeh1990PRL,Koshelev1991PC,Ong2000}, in conventional superconducting films being in the fluctuation regime \cite{Pourret2006,Behnia2016}.

Generally speaking, the Nernst signal consists of two contributions: the kinetic one and the thermodynamic one. The former is governed by the conductivity of the sample and the derivative of chemical potential of the carriers over temperature. The latter is related
to the stationary magnetization currents induced by the temperature gradient: $I_{N}^{th}=c\left(\frac{\partial m_{z}}{\partial\mathbf{T}}\right)\varDelta T$ (where $m_{z}(T)$ is the magnetization per square of the disk).  This relation was initially obtained by Obraztsov for the Hall bar geometry more than 50 years ago \cite{Obraztsov1965}. It is worthwhile to mention that this problem has been readdressed in almost every decade \cite{Smrcka1977,Jonson1984,Oji1985,Cooper1997,Qin2011,GSV2014} due to its importance for the quantum Hall effect, Nernst-Ettinghausen effect in fluctuating superconductors, anomalous thermospin effect in the low-buckled Dirac materials, etc.
The existence of magnetization currents is crucial for validity of such fundamental properties of thermomagnetic coefficients as the Onsager relations as well as the Third law of thermodynamics \cite{Obraztsov1965,SSVG2009,GSV2014}. Yet, their existence and importance for the Nernst effect often have been neglected (see for instance \cite{Dorsey1991} and discussion in \cite{Ussishkin2002PRL})  or even denied (see \cite{Reizer1,Reizer2,Reizer3}). 

The Corbino geometry offers a unique opportunity for the observation of the purely thermodynamic contribution to the Nernst effect generated exclusively by magnetization currents.  Indeed, in the regime of classically strong magnetic fields, if the chemical potential of the electron gas in the disk lies between the Landau quantization levels, the electric current does not propagate between the inner and outer edges of the disk, and one can safely neglect the kinetic part of the Nernst response.  In the same regime, in the presence of the magnetic field and temperature gradient, the contribution of the edge currents remains significant, so that the total circular current in the sample is dominated by magnetization currents.

Below  we calculate the magnetization currents of carriers characterized by
parabolic or Dirac energy dispersion relations \cite{Sharapov2004PRB} in a Corbino disk subjected
to a radial temperature gradient and a strong magnetic field $B$ applied normally to the plane. Specifically, we analyze the magnetic
field $B_{\mathrm{ind}}$ induced by these currents in the center of the disk
that can be experimentally measured e.g. by a SQUID magnetometer \cite{Barone1982,CB2004}. For the recent experimental work on the measurement of the Hall conductivity in Corbino geometry by a sensible magnetometer we refer the reader to Ref. \cite{Kapitulnik}.
We show that this field experiences pronounced unharmonic oscillations
as a function of the external field $B$. These oscillations are dominated by an interplay
of two competing factors. The background contribution to the induced magnetic field that exists
at zero temperature gradient is proportional to the second (for normal
carriers) or third (for Dirac fermions) power of the chemical potential.
At low temperatures, the chemical potential exhibits a characteristic
saw-tooth oscillatory dependence on the magnetic field that is well-known \cite{Sh1984,CM01}.
The second contribution to the magnetization, proportional to the
difference of temperatures at the inner and the outer edges of the
disk, is governed by the differential entropy per particle dependence
on the external magnetic field \cite{Pudalov}. It can be calculated knowing the
density of electronic states in the system for given temperature
and magnetic field \cite{GKV2016,GGGKPSSV2018}. The difference of the values of the induced magnetic field  $B_{\mathrm{ind}}$ measured for the opposite signs of the temperature variation between inner and outer edges of the disk is no more sensitive to the background effect and allows for extracting the contribution induced by the temperature gradient, i.e. the Nernst effect. 
The shape and the period of Nernst current oscillations
in the Corbino geometry carry a precious information on the type of
carriers and on the trajectories of topologically protected edge currents.
The universal link between the Nernst current and the induced magnetization
established in this work offers a powerful tool for the experimental
studies of transport phenomena in two-dimensional crystals.

\section*{The relation between the edge current and thermodynamic potential in the Corbino geometry}

The edge currents in Corbino geometry  can be related to the thermodynamic potential of the system basing on  very generic thermodynamic consideration.  Indeed, let us start from consideration of  a homogeneous metallic
disk of the radius $R$, placed in a thermal reservoir of temperature $T$ and subjected to the magnetic field $H$ normal to the plane of the disk. The contribution to the thermodynamic potential dependent on the induced current can be written as
\begin{equation}
\Omega_H=\frac{1}{c}\intop \mathbf{j}\left(\mathbf{r}\right)
 \mathbf{A}\left(r\right)dV
\end{equation}
where $\mathbf{A}$ is the vector potential. Consequently, the current can be expressed as
\begin{equation}
\mathbf{j}\left(\mathbf{r}\right)=\frac{c}{hS}\left(\frac{\partial\Omega_H}{\partial  \mathbf{A}}\right),
\end{equation}
with $S=\pi R^{2}$ being the area of the disk and $h$  its height. Assuming that the radius of the disk is
much larger than the magnetic length, one can choose the vector potential
in the Landau gauge, $\mathbf{A} = (0, H x)$ that yields for total current flowing through the disk
\begin{equation}
J=\frac{c}{S H}\int_{0}^{R}\frac{\partial\Omega_H}{\partial x}dx=\frac{c}{S H}\Omega_H(T).
\label{curr1}
\end{equation}
From the Fig. \ref{fig:scheme}a one can see that the current is concentrated only in the vicinity of the edge of the disk. 

Now one can represent the Corbino disk (ring) as the large disc of the radius $R_2$ from which a smaller disk of the radius $R_1$ is cut out. As a result, the total current flowing along the edges is given by the difference between the outer and internal edge currents. Both currents are defined by the same Eq. (\ref{curr1}), taken with different areas of the disk.  Accounting for the temperature difference between the edges, one can finally obtain
\begin{equation}
J_{tot}=\frac{c}{H}\left[\frac{\varOmega(T_{2})}{S_{2}}-\frac{\varOmega(T_{1})}{S_{1}}\right].
\end{equation}

The derivation above is based on classical arguments that may seem contradictory to the quantum nature of Landau diamagnetism. This is why, in the following section we will reproduce the final expression for the current (\ref{curr1}) in the framework of a quantum mechanical approach.
\section*{The microscopic approach to calculation of edge currents}

\subsection*{The eigenvalue problem}

Let us consider now the Corbino disk with the inner edge radius $R_{1}$
and the outer edge radius $R_{2}$ subjected to the magnetic field $H$
applied normally to the disk plane in microscopic approach. We are interested here in the
regime of classically strong magnetic fields, where the energy separation between
the neighbouring Landau levels exceeds their broadening, yet remaining
small with respect to the Fermi energy: $\mbox{max}\left\{ T,\varGamma\right\} \ll\hbar\omega_{c}\ll E_{F}$, where $T$ is temperature, $\varGamma$ is the Dingle temperature,
$\omega_{c}$ is the cyclotron
frequency, $E_{F}$ is the Fermi energy. In what concerns the requirements
to the disk geometry, we assume that $R_{1},R_{2},R_{2}-R_{1}\gg l_{B}$,
where the magnetic length is $l_{B}=\sqrt{\hbar c/|e|H}$.

\begin{figure}[!t]%[tbhp]
\centering
\includegraphics[width=0.95\linewidth]{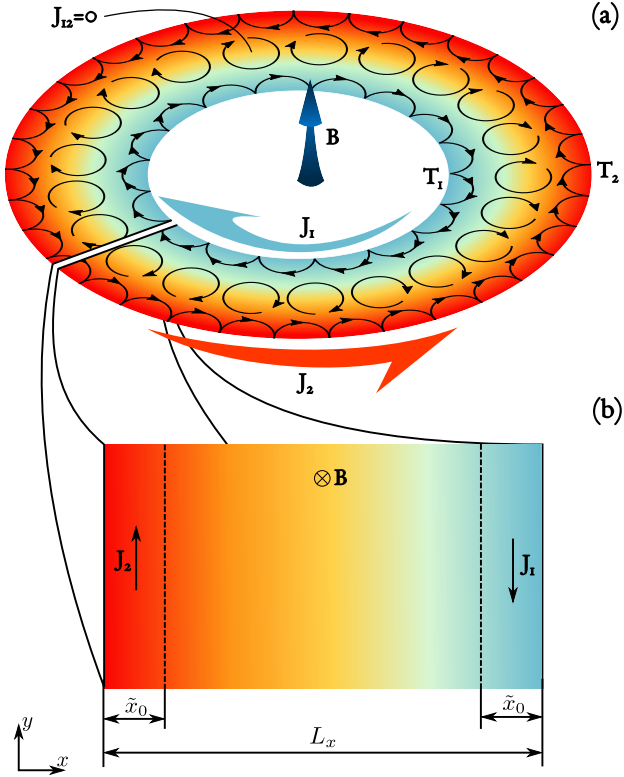}
\caption{a). The schematic showing the edge currents flowing in a Corbino disk subjected to an external magnetic field normal to its plane and to a radial temperature gradient. b). The schematic showing the edge currents flowing in a conducting strip subjected to an external magnetic field normal to its plane.}
\label{fig:scheme}
\end{figure}

We use here the thermodynamic approach to the Nernst effect developed in Refs.~\cite{Obraztsov1965,LKV2011PRL}. Namely, we
describe the system by the Gibbs thermodynamic potential
\begin{equation}
\Omega=-2kT\sum_{\alpha} \ln \left[1+e^{\frac{\mu-\varepsilon_{\alpha}}{kT}}\right],
\label{eq:gibbs}
\end{equation}
where $\varepsilon_{\alpha}$ are the eigenvalues, and the summation
is performed over the complete set of the quantum numbers $\left\{ \alpha\right\} $,
$\mu\left(T\right)$ is the chemical potential of the electron gas,
coinciding with the Fermi energy at zero temperature. The spin degeneracy of the electron gas under study
is postulated, that results in the appearance of the factor 2 in Eq.~(\ref{eq:gibbs}).
In the case corresponding to the 2D gas of free electrons (2DEG) subjected to a
magnetic field, the electronic Hamiltonian has the familiar Landau form (the specifics
of carriers having a Dirac dispersion will be discussed later):
\begin{equation}
\hat{H}=-\frac{\hbar^{2}}{2m}\frac{d^{2}}{dx^{2}}+\frac{m\omega_{c}^{2}}{2}\left(x-x_{0}\right)^{2}\label{hamilton}
\end{equation}
with $\omega_{c}= |e| H /\left(mc\right)$. The set of quantum numbers is $\alpha=\left\{ x_{0},n\right\} $,
where $x_{0}=l_{B}^{2}k_{y}$ is the $x$-coordinate of the center of the electron cyclotron
orbit, $k_{y}$ is the tangential component of the electron momentum (see the schematic
in Figure \ref{fig:scheme}a), $n$ is the index of the energy quantization level in the
potential induced by the magnetic field (with the minimum at the point $x_{0}$).

The rule for summation over eigenvalues in Eq. (\ref{eq:gibbs}) takes a
form
\begin{equation}
\sum_{\alpha}...=\frac{|e|H}{c}\frac{L_{y}}{2\pi\hbar}\int_{-\infty}^{\infty}dx_{0}\sum_{n=0}^{\infty}...,\label{rule}
\end{equation}
where $L_{y}$ is the linear dimension of the system along the edge.

The Shr\"{o}dinger equation with the Hamiltonian (\ref{hamilton})
and the specific boundary conditions $W_{\alpha}\left(0\right)=W_{\alpha}\left(L_{x}\right)=0$
(Teller's model \cite{Teller1931}) determine the spectrum $\varepsilon_{\alpha}$
and the set of eigenfunctions:
\begin{equation}
\hat{H}W_{\alpha}\left(x\right)=\varepsilon_{\alpha}W_{\alpha}\left(x\right).\label{shrod}
\end{equation}
The latter turn out to be the Weber functions $W_{\alpha}\left(x\right)$ Ref.~\cite{Bateman.book},
while the electron eigenenergies in the vicinity of the edge $x=0$
can be approximated by:
\begin{equation}
\varepsilon_{\alpha}\!=
\hbar \omega_c\!
\begin{cases}
\begin{array}{@{}l@{}r@{}}
%\hbar\omega_{c}
\left(n\!+\! \frac{1}{2}\right)\!+\! %\hbar\omega_{c}
\frac{2^{n}}{\sqrt{\pi}n!}\!\left(\frac{x_{0}}{l_{B}}\right)\!^{2n+1}
\!\exp\!\left[\!-\frac{x_{0}^{2}}{l_{B}^{2}}\right]\!, & \; x_{0}\!\gg l_{B}\\
2 %\hbar\omega_{c}
\left(n+ \frac{3}{4}\right)-2%\hbar\omega_{c}
\frac{\left(2n+1\right)\Gamma\left(n+1/2\right)}{\pi n!}\left(\frac{x_{0}}{l_{B}}\right), & \; x_{0}\lesssim\! l_{B}
\end{array}\end{cases}\label{eigen1}
\end{equation}
(we note that 
the similar expressions were obtained in Ref. \cite{Hajdu1974ZP}, while some errors in the coefficients and the erroneous factor of ``2'' in the exponential function are present in that work.)
The
upper line in (\ref{eigen1}) corresponds to the cyclotron orbits
centered far from the edges ($x_{0}\gg l_{B}$). The energy spectrum
for these states coincides with the Landau  one with an exponential
accuracy. The lower line describes the energy spectrum for the states
whose orbits are centered close to the border ($x_{0}\lesssim l_{B}$).
The doubling of the cyclotron frequency that appears in the first
term is due to the supplementary quantum confinement of carriers in
a half-parabolic potential that appears due to their reflection from
the boundary.

\subsection*{The edge currents calculated from the first principles}
%\begin{figure}%[tbhp]
%\centering
%\includegraphics[width=1.0\linewidth]{currents.eps}
%\caption{The schematic showing the edge currents flowing in a %conducting strip subjected to an external magnetic field normal %to its plane.}
%\label{fig:strip}
%\end{figure}
We consider a macroscopic Corbino disk and assume that the curvature of the edges can be safely neglected on the length scale of the cyclotron orbits (see Fig. \ref{fig:scheme}b). In this case, one can calculate the edge currents starting from the exact quantum mechanical expression for the charge flow in a pure quantum state $\alpha$ \cite{LLIII,Hajdu1974ZP}:
\begin{equation}
j_{y\alpha}\left(x,x_{0}\right)=-\frac{|e|\omega_{c}}{L_{y}}\left(x-x_{0}\right)W_{\alpha}^{2}\left(x\right).\label{flow}
\end{equation}
The full current $J_{\mathrm{tot}}$ is obtained by summing $j_{y\alpha}$ over all eigenstates
$\left\{ \alpha\right\} $ of the problem, accounting for the occupation
numbers $f\left(\varepsilon_{\alpha}\right)=\left[\exp\left(\left(\varepsilon_{\alpha}-\mu\right)/kT\right)+1\right]^{-1}$
and integrating over the width of the disk:
\begin{equation}
\begin{split}
& J_{\mathrm{tot}} =\int_{0}^{L_{x}}\sum_{\alpha}j_{y\alpha}\left(x,x_{0}\right)f\left(\varepsilon_{\alpha}\right)dx\\
%& =-m\frac{|e|\omega_{c}^{2}}{\pi\hbar}\sum_{n=0}^{\infty}\int\limits_{-\infty}^{\infty} dx_{0}f\left[\varepsilon_{n}\left(x_{0}\right),T\right]
& = -m\frac{|e|\omega_{c}^{2}}{\pi\hbar}
\!\sum_{n=0}^{\infty}\!\int\limits_{-\infty}^{\infty} \!\!\!dx_{0}f\left[\varepsilon_{n}\left(x_{0}\right),T\right]
\!\!\!\int\limits_{0}^{L_{x}}\!\!dx\left(x\!-\!x_{0}\right)W_{\alpha}^{2}\left(x\right).\label{current}
\end{split}
\end{equation}
One can relate the integral over $x$ in Eq. (\ref{current}) to the
derivative of the eigenenergy over $x_{0}$ employing the Feynman
theorem \cite{LLIII}:
\begin{equation}
\int_{0}^{L_{x}}dx\left(x-x_{0}\right)W_{\alpha}^{2}\left(x-x_{0}\right)=\frac{1}{m\omega_{c}^{2}}\frac{\partial\varepsilon_{\alpha}}{\partial x_{0}},\label{eq:feynman}
\end{equation}
that results in
\begin{equation}
J_{\mathrm{tot}}\!\!=\!-\frac{|e|}{\pi\hbar}\!\sum_{n=0}^{\infty}\int\limits_{-\infty}^{\infty}\!\!\!\!\frac{d}{dx_{0}} \ln\!\left[1+\exp\left(\!\!\frac{\mu\left(T\right)-\varepsilon_{n}\left(x_{0}\right)}{kT}\!\right)\!\right]\!\!dx_{0}.\label{current-1}
\end{equation}
We underline that far from the edges of the disk
the electron energy levels (\ref{eigen1}) coincide with the Landau levels with an exponential accuracy,
i.e. in this domain $\widetilde{x}_{0} \lesssim x\lesssim L_{x}-\widetilde{x}_{0}$
the derivative $\partial\varepsilon_{\alpha}/\partial x_{0}$=0. The
value $\widetilde{x}_{0}$ can be estimated by imposing the phenomenological
requirement\cite{Hajdu1974ZP}:
\[
\left(\frac{\partial\varepsilon_{\alpha}}{\partial x_{0}}\right)_{x_{0}=\widetilde{x}_{0}}=0.
\]
One can see from Eq. (\ref{eigen1}) that $\widetilde{x}_{0}\sim l_{B}\sqrt{2n+1}$
, which is nothing but the radius of the cyclotron orbit at the $n$-th
Landau level. Having this in mind, the integration in
(\ref{current-1}) can be restricted to the vicinity of the edges of the
sample: $]-\infty,\widetilde{x}_{0}]$, $[L_x-\widetilde{x}_{0},\infty[$.
The contribution to the current from the bulk region tends to
zero (see Fig.~\ref{fig:scheme}).

\subsection*{The Corbino disk with differently heated edges}

In order to study the Nernst effect we assume that the inner (outer)
edge of the disk is kept at equilibrium with the thermal bath of the temperature
$T_{1}\left(T_{2}\right).$ We assume that the temperature 
gradient is small enough, so that on the scale of the order
of $\widetilde{x}_{0}$ it can be neglected.
In this case the full circular current  is determined by the difference of two edge currents:
\begin{equation}
J_{\mathrm{tot}}=J \left(T_{1}\right)-J\left(T_{2}\right),
\label{difference}
\end{equation}
where
\begin{equation}
J\left(T\right)=-\frac{|e| k T}{\pi\hbar}\!\sum_{n=0}^{\infty}\ln\!\left[1+\exp\left(\!\!\frac{\mu\left(T\right)-\varepsilon_{n}\left(\tilde{x_{0}}\left(n\right)\right)}{kT}\!\!\right)\!\right]. \label{eq:finalgen}
\end{equation}
Since the sum in (\ref{eq:finalgen}) is determined by its upper limit one 
can use the expression for $\varepsilon_{n}\left(\tilde{x_{0}}\left(n\right)\right)$
from the upper line of Eq. (\ref{eigen1}). Neglecting the exponentially
small second term, we reproduce the result discussed in \cite{Mineev2007PRB}:
\begin{equation}
J\left(T\right)\approx-\frac{|e| k T}{\pi\hbar}\!\sum_{n=0}^{\infty}\ln\!\left[1\!+\!\exp\!\left(\!\!\frac{\mu\left(T\right)\!-\!\hbar\omega_{c}\left(n\!+\!1/2\right)}{kT}\!\right)\!\right].\label{final current}
\end{equation}
The Eqs. (\ref{difference})-(\ref{final current}) describe
the total current induced by the external magnetic field in the Corbino
geometry. The chemical potential $\mu\left(B,T,\rho\right)$ depends
on the magnetic field, temperature, and the carrier concentration $\rho$. Comparing
Eq. (\ref{final current}) with the thermodynamic potential calculated
for the Landau energy spectrum (see Eqs. (\ref{eq:gibbs})- (\ref{rule}))
\begin{equation}
\begin{split}
& \Omega_{L}\left(T\right)=-2kT\frac{|e|H}{c}\frac{S}{2\pi\hbar} \\
& \times \sum_{n=0}^{\infty}\ln\left[1+\exp\left(\frac{\mu\left(T\right)-\hbar\omega_{c}\left(n+1/2\right)}{kT}\right)\right],\label{eq:omegaL}
\end{split}
\end{equation}
 one finds the universal relation
which was first derived by Obraztsov in \cite{Obraztsov1965}:
\begin{equation}
\begin{split}
J\left(T,H,\mu\right) =\frac{c}{HS}\Omega_{L}\left(T,H,\mu\right).
\end{split}
\label{eq:currentpotent}
\end{equation}
Let us stress that the sign in Eq.~(\ref{eq:currentpotent}) is the matter of convention: in the chosen form it corresponds the direction of the current flowing along the internal edge of the ring.

The problem of calculation of the Gibbs potential in the presence of a homogeneous magnetic
field was considered long  ago in relation to the de Haas - van Alphen
oscillations. The corresponding expression can be easily obtained from
(\ref{eq:omegaL}). In the limit of low temperatures $kT\ll\mu\left(T\right)$,
the exponential term in the argument of the logarithmic function strongly exceeds unity. 
The expression for the current thus reduces to
\begin{equation}
J^{\mathrm{2DEG}}(T,\mu)=-\frac{|e|}{\pi\hbar^{2}}\frac{\mu^{2}\left(T\right)}{2\omega_{c}}.
\label{eq:musquare}
\end{equation}

In the case of graphene characterised by the linear dispersion of Dirac carriers, the Landau quantization leads to the appearance of a non-equidistant energy spectrum ($E_n = \pm \sqrt{2 n \hbar |e| B v_F^2/c }$), in which case the summation in Eq.(\ref{eq:finalgen}) results in
\begin{equation}
J^{\mathrm{gr}}(T,\mu)=-\frac{c}{H} \frac{|\mu (T)|^{3}}{3 \pi \hbar^2 v_F^2},
\label{eq:muscub}
\end{equation}
where $v_F$ is the Fermi velocity.

%Introducing the graphene cyclotron frequency $\omega_c = |eB| v_F^2/(c |\mu|)$ one can rewrite

\section*{The Nernst oscillations in 2DEG and graphene}
\subsection*{Oscillations of the edge currents}

The local chemical potential $\mu (T_{1,2},H)$ at the edge of the disk as a function of the temperature and 
magnetic field is defined by the equation
\begin{equation}
\label{mu-equation}
\frac{1}{S}\left. \frac{\partial \Omega_{L}\left(T,H,\mu\right) }{\partial \mu} \right|_{H,T_{1,2}}  = - \rho.
\end{equation}
Here $\rho$ is the concentration of carriers.

At low temperatures, in a  system with a fixed carrier density
the chemical potential can be presented as 
$\mu(T,H) = E_{F} + \widetilde{\mu}(T,H)$, where the oscillating part of
the chemical potential  $\widetilde{\mu}$ 
is given by  Eq.~(\ref{mu-equation}) 
\cite{CM01,LKV2011PRL,Sharapov2004PRB}.
This equation that is implicit for $\widetilde{\mu}$ can be solved analytically only at 
$T=0$ \cite{Grigoriev2001JETP}, while at finite temperatures one needs to resort to the numerical analysis. In the case of weak smearing of the
Landau levels $ \widetilde{\mu}$  turns to be  of the order of $\hbar \omega_c$. 
However, if $\Gamma$ approaches $\hbar \omega_c/(2\pi)$, the value of 
$ \widetilde{\mu}$ exponentially decays \cite{CM01}. In this case the approximate
expression for $ \widetilde{\mu}$   both for a 2DEG of carriers having a parabolic dispersion and for Dirac fermions in graphene can be written:
\begin{equation}
\label{mu-osc}
\begin{split}
& \widetilde{\mu} =  \\
& - \!\frac{ \hbar \omega_c}{\pi}\!\!  \sum\limits_{l=1}^\infty \!\!\frac{\psi(l \lambda)}{l} 
\sin\!\! \left[\!2 \pi l\! \left(\!\!\frac{c \mathcal{S}(E_F)}{2 \pi e \hbar B} \!+\! \frac{1}{2} \! + \!\beta \!\right)\!\!\right]\!\!
\exp\!\left(\!-\frac{2\pi l\Gamma}{\hbar \omega_{c}}\right),
\end{split}
\end{equation}
where 
\begin{equation}
\psi (z) =
\frac{z}{\sinh z}, \qquad
\lambda =\frac{2\pi^{2} k T}{\hbar \omega_{c}}
\end{equation}
is the temperature factor, $\mathcal{S}(E_F)$ is the electron cyclotron orbit area in the momentum space, $\beta$ is the
topological part of the Berry phase.
In the case of a 2DEG characterized by a parabolic dispersion of charge carriers, $\epsilon = p^2/(2 m)$, the electron orbit area
is $\mathcal{S}(E_F) = 2 \pi m E_F$, and the trivial phase, $\beta = 0$.
In its turn, for the massless Dirac fermions, 
$\epsilon = \pm v_F p$, the area is
$\mathcal{S}(E_F) = \pi E_F^2/v_F^2$, while the cyclotron frequency  
depends on the Fermi energy $\omega_c = v_F^2 |e | H/(c|E_F|)$. In contrast to the case of a 2DEG with a parabolic dispersion, for Dirac fermions
the phase becomes nontrivial,
$\beta = 1/2$. All above is valid for the range of classically strong magnetic
fields, $\hbar \omega_c \ll E_F$ (we assume that $E_F > 0$).

Substituting  Eq.~(\ref{mu-osc}) to Eqs.~(\ref{eq:musquare}) and (\ref{eq:muscub})
one can find explicitly the magnetic field dependence of the edge currents:
\begin{equation}
J\left(T,H\right)=-\frac{|e|}{\pi\hbar^{2}}  \frac{E_{F}^{2}}{\omega_{c}} 
\left[ \eta
+ \frac{ \widetilde{\mu}  }{E_F} \right]
\end{equation}
with $\eta = 1/2$ for 2DEG and $\eta = 1/3$ for the Dirac electrons in graphene, respectively.

Applying the above expression for the edge currents flowing in the Corbino disk with differently heated inner and outer edges one can find the sum of two edge currents as
\begin{equation}
J_{tot} \left(T_{1},T_{2}\right)=\frac{|e|}{\pi\hbar^{2}}
\frac{E_{F}}{\omega_{c}}\left[\widetilde{\mu}\left(T_{2},H\right)-
\widetilde{\mu} \left(T_{1},H\right)\right].
\end{equation}

\begin{figure}[!ht]%[tbhp]
\centering
\includegraphics[width=8.65cm]{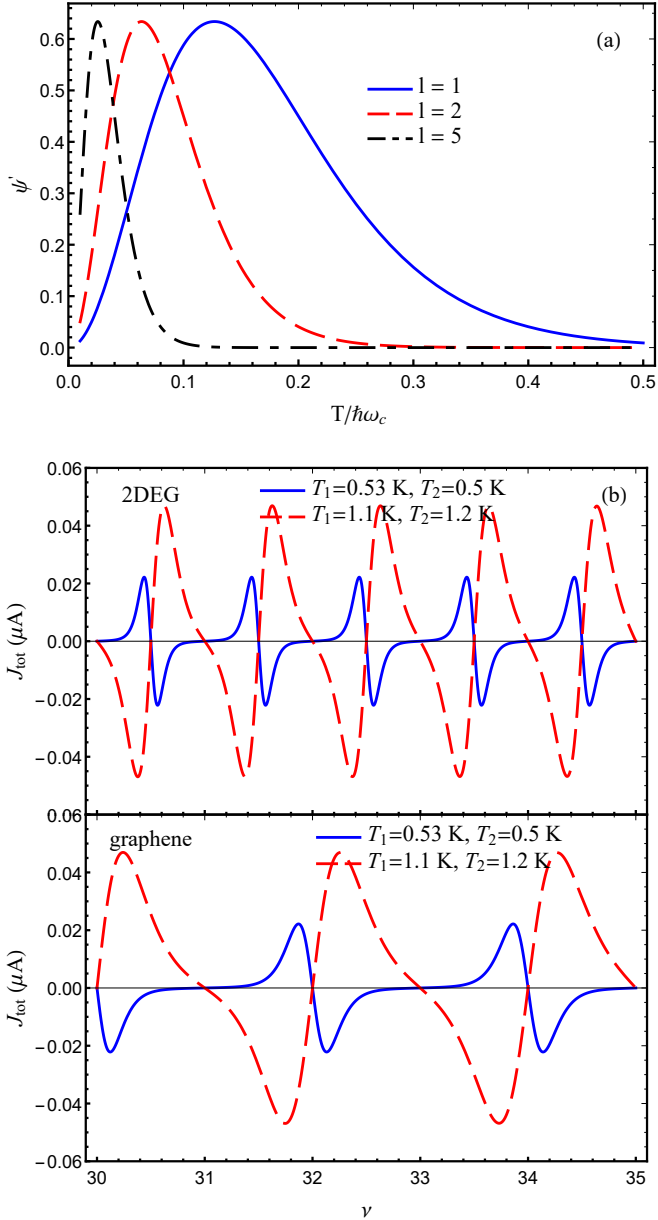}
\caption{(a) The dimensionless amplitude factor (\ref{thermal}) plotted as a function of temperature $T$ measured in the units of $\hbar \omega_c$
for three different values of $l$. (b) The sum of two edge currents $J_{tot}$ in $\mu A$ as a function the Landau filling factor $\nu$ that is introduced for the cases of 2DEG and graphene in 
the body of the paper. 
The Fermi energy is assumed to be
$E_F = 500 \, \mbox{K}$ and the level broadening $\Gamma = 0.5 \, \mbox{K}$. 
The cyclotron energy $\hbar \omega_c = E_f/\nu$.
Note that $T_1 > T_2$ for the blue curves and $T_1 < T_2$ for the red ones. 
The direction of the temperature gradient strongly affects the shape of 
the oscillations. }
\label{fig:current}
\end{figure}

%\begin{equation*}
%\begin{split}
%&=\frac{|e|}{\pi^{2}\hbar}E_{F}\sum_{l=1}^{\infty}\frac{\left(-1\right)^{l+1}}{l}\left[\frac{\lambda_{l}\left(T_{in}\right)}{sinh\lambda_{l}\left(T_{in}\ri%ght)}-\frac{\lambda_{l}\left(T_{out}\right)}{\sinh\lambda_{l}\left(T_{out}\right)}\right]\\
%& \times
%sin\left(\frac{2\pi lE_{F}}{\omega_{c}}\right)e^{-\left(-\frac{2\pi l\Gamma}{\omega_{c}}\right)}.
%\end{split}
%\end{equation*}
In the case of a relatively small temperature difference $\Delta T=T_1 - T_2 \ll T_1$ one can expand $\widetilde{\mu}$
and obtain the explicit dependence of the oscillating total current on the magnetic field:
\begin{equation}
\begin{split}
J_{tot} &\left(T,\Delta T \right) = \frac{|e|E_{F}}{\pi^{2}\hbar}\left(\frac{\Delta T}{T}\right)
\sum_{l=1}^{\infty} 
\frac{\psi^{\prime}( l \lambda)}{l} \\
&\qquad
\times
\sin \left[2 \pi l \left(\frac{c \mathcal{S}(E_F)}{2 \pi e \hbar B} \!+\! \frac{1}{2} \! + \!\beta \right)\right]
\exp\left(-\frac{2\pi l\Gamma}{\hbar \omega_{c}}\right).
\end{split}
\end{equation}
The amplitude factor
\begin{equation}
\psi^{\prime}(l \lambda)=
\frac{\lambda l \left[\lambda l \coth (\lambda l) -1\right]}{\sinh (\lambda l) } =
\begin{cases}
\frac{\lambda l}{3} \left[ 1 - \frac{7}{30} (\lambda l)^2\right], & \lambda l \ll 1, \\
2 \lambda l \exp(- \lambda l), &  \lambda l \gg 1 %
\end{cases}
\label{thermal}
\end{equation}
is presented in Fig.~\ref{fig:current}~a. In contrast to the conventional factor
$\psi(l \lambda)$, it is a nonmonotonic function of temperature.

Note that the same function $\psi^{\prime}(l \lambda)$ appears in the expression for the
oscillating part of the Seebeck coefficient in an electron gas subjected to a strong magnetic field \cite{Varlamov1989PLA}.

The total edge current $J_{tot} \left(T_1, T_2, \nu \right) $
as a function of the filling factor of a Landau level
\begin{equation}
\label{filling-standard}
\nu  = \frac{ \pi \hbar c \rho}{eB} =  \begin{cases}
\frac{ E_F}{\hbar \omega_c}, & \mbox{2DEG} \\
\frac{c E_F^2}{\hbar e H v_F^2}, &  \mbox{graphene}%
\end{cases}     
\end{equation}
is plotted in Fig.~\ref{fig:current}~b.
One can see that the period of oscillations for graphene is twice larger due 
to the valley degeneracy. The phase of oscillations for graphene is shifted 
with respect to 2DEG.
The sharp features correspond to the Fermi level crossing by the Landau levels.
Note that the shape of obtained current oscillations shown in Fig.~\ref{fig:current}~b resembles one of the thermoelectric power coefficient for the 2DEG calculated in
\cite{Varlamov1989PLA}.

We recall that all our calculations are performed within the assumption of the smallness of the magnitude of the $\widetilde{\mu} \ll \hbar \omega_c$, that implies a noticeable
smearing of Landau levels ($\Gamma/k + T \sim 1.5 \, \mbox{K} 
\sim  \hbar \omega_c /(2\pi k) \sim 2 \, \mbox{K}  $). It is important to note that this experimentally relevant limit allows to demonstrate 
the existence of unusual Nernst oscillations which appear due to the imbalance of magnetization currents in the Corbino geometry.

\subsection*{The induced magnetic field and its oscillations in the Corbino geometry}

\begin{figure}[!t]%[tbh]
\centering
\includegraphics[width=8.65cm]{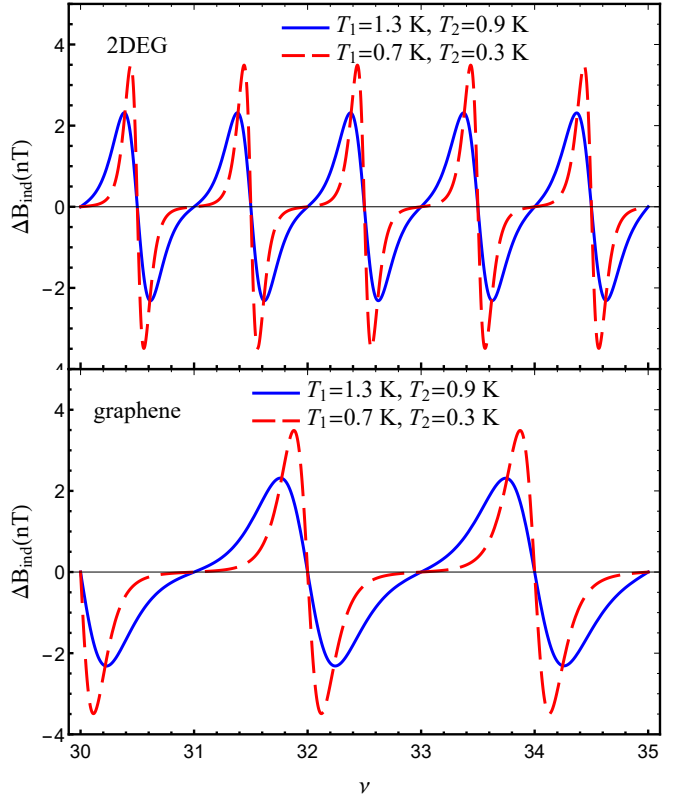}
\caption{The contribution to the induced magnetic field $\Delta B_{ind}$ at the center of the Corbino disk
that is induced by a temperature gradient
in $nT$ plotted as a function of the filling factor $\nu$. 
The upped panel is describing the 2DEG characterised by a parabolic dispersion of charge carriers while the lower panel corresponds to the case of graphene characterised by the linear dispersion of charge carriers. 
The blue curves are calculated with $T_1 = 1.3 K $,  $T_2 = 0.9 K $
and the red curves correspond to $T_1 = 0.7 K $,  $T_2 = 0.3 K $.
The other parameters used in this calculation are $R_1 = 100 \mu m$, 
$R_2 = 110 \mu m$, $E_F = 500 \, \mbox{K}$, $\Gamma = 0.5 \, \mbox{K}$. 
}
\label{fig:B-osc}
\end{figure}

The circular electric currents $J(T_{1,2})$ along the edges
of the disk lead to the induction of the magnetic field 
in the center of the disk
$B_{ind}(T_1,T_2) = B_1 + B_2$ with 
$B_{1,2}=\pm2\pi J(T_{1,2})/cR_{1,2}$. 
This field constitutes a diamagnetic response of the ring generated by the persistent
currents that have a purely thermodynamic nature
\begin{equation}
\label{Bind}
B_{ind}(T_1,T_2) = \eta \frac{|e|E_{F}}{\hbar c}\left(\frac{E_{F}}{\hbar\omega_{c}}\right)\left(\frac{1}{R_{1}}-\frac{1}{R_{2}}\right) + B_{osc}.
\end{equation}
The first term in Eq.~(\ref{Bind}) monotonously 
decreases with the increase of the external magnetic field as a result of the reduction of the magnitude of the edge currents.
The oscillating part $B_{osc}$ of the induced magnetic field for the specific cases
of carriers with parabolic and linear dispersions is given by
\[
\begin{split} & B_{osc}=\frac{|e|E_{F}}{2 \pi \hbar c}\sum_{l=1}^{\infty}\frac{1}{l}\left[\frac{\psi\left[l\lambda\left(T_{1}\right)\right]}{R_{1}}-\frac{\psi\left[l\lambda\left(T_{2}\right)\right]}{R_{2}}\right]\\
 & \times\sin\left[2\pi l\left(\frac{c \mathcal{S}(E_{F})}{2\pi e\hbar B}\!+\!\frac{1}{2}\!+\!\beta\right)\right]\exp\left(-\frac{2\pi l\Gamma}{\hbar\omega_{c}}\right).
\end{split}
\]
In order to exclude the background part of $B_{ind}$ that is independent of the temperature gradient 
one can study the difference of the induced fields $\Delta B_{ind}(T_1,T_2) =  B_{ind}(T_1,T_2) -  B_{ind}(T_2,T_1) $. 
The dependence of $\Delta B_{ind}(T_1,T_2)$ on the filling factor is shown in Fig.~\ref{fig:B-osc}. 
%One can see that the mean value of the induced field increases with the %increase of the Landau filling factor. 
The phase and magnitude of the oscillatory features corresponding to the resonances of Landau and Fermi levels is strongly dependent on the temperature gradient in the studied sample.
%Here  we used
%\begin{equation}
%\label{Ef2n}
%E_F = \frac{\pi n \hbar^2}{m} = 2.4 \times 10^{-4} \frac{m_e}{m} x [\mbox{eV}] = 2.8 \frac{m_e}{m} x \mbox{[K]},
%\end{equation}
%where spin degeneracy is included. Yet, accounting for the spin degeneracy we add a factor of 2 in the denominator of (\ref{filling-standard}), so that we arrive at the following expression for the filling factor
%\begin{equation}
%\nu = \frac{\pi \hbar c n}{|eB|} = \frac{E_F}{\hbar \omega_c}.
%\end{equation}
%All above is for 2DEG. 
The oscillations depend on the temperature difference at the edges of the disk. Dirac fermions and 2D electrons with a parabolic spectrum are characterized by oscillations of different phase and frequency. We predict qualitatively different power dependencies of the magnitude of Nernst signal on the chemical potential for normal and Dirac carriers. 

\section*{Conclusions}

 We have demonstrated that the Corbino geometry offers a precious opportunity for the observation of the specific Nernst effect having a purely thermodynamic nature. The effect is caused by the imbalance of magnetization currents flowing along the inner and outer edges of the Corbino disk maintained at different temperatures. We  demonstrate  that the experimentally observable manifestation of this thermodynamic Nernst effect consists in the appearance of the specific oscillations of the  magnetic field measured in the center of the disk as a function of the external field. The measurement should be done in the presence of a temperature gradient between the inner and outer edges of the disk. Subtracting the values of the magnetic field measured in the center of the disk for opposite signs of the temperature gradient one should be able to extract the specific contribution of the magnetisation currents to the Nernst signal.
 
 We have developed the microscopic model describing such oscillatory diamagnetic response of the Corbino disk made of a normal metal and of graphene in the presence of the radial temperature gradient. The total current exhibits oscillations corresponding to the resonances of Fermi and Landau levels in the disk. The value and the direction of the radial temperature gradient in the sample strongly affect the magnitude and the shape of the oscillations in the dependence of the induced magnetic field on the Landau filling factor. An experimental study of such diamagnetic oscillations in the center of the Corbino disk would allow for the high precision measurement of the Nernst effect that is expected to be of strongly different magnitude in graphene and in normal metals. Such a study would also shed light on the contribution of the diamagnetic currents to the Nernst effect that has been a subject of debate for many years. 

\acknow{A.A.V. and S.G.Sh. acknowledge the hospitality of the Westlake University, where this work was started and mainly accomplished, the support of EC for the RISE Project CoExAN GA644076. They also are grateful to C.~Goupil and A.~Chudnovskiy for illuminating discussion.
A.A.V. acknowledges a support by  European Union's Horizon 2020  research and innovation program under the grant agreement n 731976 (MAGENTA).
S.G.Sh. acknowledges a support by the National Academy of
Sciences of Ukraine (project No. 0117U000236) and by its Program of
Fundamental Research of the Department of Physics and Astronomy (project No.
0117U000240). P.S.G. acknowledges Saint-Petersburg State University for a research grant ID 40847559}

\showacknow{} % Display the acknowledgments section

% Bibliography
%\bibliography{pnas-biblio}
\bibliography{mainbib}

\iffalse

\fi

\end{document}